# Compression and Encryption of Search Survey Gamma Spectra using Compressive Sensing

Alexander Heifetz, Sasan Bakhtiari, and Apostolos C. Raptis, *Life Member, IEEE*

***Abstract*—We have investigated the application of Compressive Sensing (CS) computational method to simultaneous compression and encryption of gamma spectra measured with NaI(Tl) detector during wide area search survey applications. Our numerical experiments have demonstrated secure encryption and nearly lossless recovery of gamma spectra coded and decoded with CS routines.**

## I. Introduction

DETECTION and identification of nuclear sources in urban search surveys, or verification of the absence of sources in wide area screenings, constitutes an important problem for national security. In a source search and wide area screening, gamma spectra are continuously measured in air in short acquisition time intervals (e.g. one second) with a moving nuclear detector-spectrometer [1]. Efficient wide area searches and screenings require a network of coordinated nuclear detectors. In such network, nuclear detectors would transmit the spectra measured every second, in real time, over wireless communication channels. Although the system of wireless nuclear detector network should increase the efficiency of source search compared to the case when a single nuclear detector is used, information transmission over wireless channels leaves the network vulnerable for a cyberattack. In addition, wireless channels may have limited communication bandwidth. Therefore, it is advantageous to compress and encrypt nuclear spectra for bandwidth-efficient and secure transmission of data. The downside of adding data compression and encryption layer is that slows down network operation. In nuclear search, measured spectrum is sent from a mobile detector to the central processor every second. After the compressed and encrypted signal is received and decoded, it needs to be analyzed in real-time to detect the presence or absence of a source of interest. Therefore, there is a need to develop high performance data encryption and compression algorithms for networks of wireless nuclear detectors. Compressive Sensing (CS) is a computational technique [2-4] which has been shown to enable simultaneous compression and encryption of data [5,6]. In this paper, we are investigating performance of CS-based compression and encryption algorithms with eventual goal of demonstrating advantages of CS-based methods over the existing state-of-the-art data compression and encryption algorithms [7,8].

Manuscript received September 16, 2014. This work was supported by the Argonne National Laboratory LDRD 2011-027-N0 "Data Compression and Encryption for Cybersecurity Applications."

A. Heifetz is with the Nuclear Engineering Division, Argonne National Laboratory, Lemont, IL 60439 USA (phone: 630-252-4429; fax: 630-252-3250; e-mail: aheifetz@gov).

S. Bakhtiari is with the Nuclear Engineering Division, Argonne National Laboratory, Lemont, IL 60439 USA (phone: 630-252-8982; fax:630-252-3250; e-mail: bakhtiari@anl.gov).

## II. Motivation for Application of Compressive Sensing

### *A. Overview of Compressive Sensing*

CS is a computational technique that enables efficient reconstruction of information from significantly undersampled number (below Shannon-Nyquist limit) of linear measurements by taking advantage of the data sparsity [2-9]. A signal $\mathbf{x} \in R^n$ is sparse in the basis $\mathbf{\Phi} = \{\boldsymbol{\varphi}_1, ... \boldsymbol{\varphi}_n\}$ if in the representation $\mathbf{x} = \mathbf{\Phi} \cdot \boldsymbol{\alpha} = \alpha_1 \boldsymbol{\varphi}_1 + ... + \alpha_n \boldsymbol{\varphi}_n$, most entries of the coefficient vector $\boldsymbol{\alpha}$ are zero or close to zero. In general, one can find a basis for most naturally occurring signals where the signals are sparse. In CS, measurements (samplings of the signal which results in compression) are obtained by linearly projecting the signal $\mathbf{x}$ on a set of measurement vectors $\mathbf{A} = \{\mathbf{a}_1, ... \mathbf{a}_n\}$, where $\mathbf{a}_i \in R^m$ and $m \le n$. In matrix form, the measured (compressed) vector $\mathbf{y} \in R^m$ is $\mathbf{y} = \mathbf{A} \cdot \mathbf{x}$. The requirements for CS applicability is that $\mathbf{x}$ is sparse in $\mathbf{\Phi}$ and measurement vectors in $\mathbf{A}$ are non-adaptive (independent of $\mathbf{x}$) and incoherent with $\mathbf{\Phi}$ [9]. Recovery step involves solving convex optimization problem with $l_1$-norm minimization

$$\min \| \boldsymbol{\alpha} \|_{l_1} \ s.t. \| \mathbf{y} - \mathbf{A\Phi}\alpha \|_{l_2} \le \varepsilon , \qquad (1)$$

where $\varepsilon$ is the tolerance. This problem is known as LASSO or Basis Pursuit denoising. Several methods exist for solving the inverse problem in Eq. (1) [10-13].

### *B. Potential Advantages of Compressive Sensing Approach*

CS method carries and additional advantage in that the compressively sampled data is automatically encrypted. In the CS scheme is the measurement matrix **A** is both the compression and encryption key. The measurement matrix is known to both the sender and the receiver. Without knowledge of the measurement matrix, the received signal cannot be decrypted by an unauthorized listener, such as an eavesdropper intercepting the signal. Blind reconstruction of the signal is a NP-hard problem, which provides a measure of security against eavesdropper via the computational infeasibility of reconstruction.

CS is particularly suitable for data compression in distributed sensor networks, where computationally lightweight coding is performed by a sensor's on-board processor, while computationally intensive reconstruction is performed by the network's receiver processor. This is advantageous for a network of hand-held nuclear detectors with limited battery power. Since CS combines data compression and encryption in one operation, this method has potential advantages in performance speed over the tradition protocols that involve sequential compression and encryption operations.

## III. Overview of Nuclear Search Data

In this paper, we investigated approaches to encryption and compression of spectral data collected with Thallium-doped Sodium Iodide (NaI(Tl)) scintillation detector-spectrometer. NaI(Tl) nuclear detector is most frequently used in search applications because of its fast response and relatively low cost. A modern NaI detector measures gamma radiation in the energy range 0 to 3MeV using 1024 spectral channels. Thus, the measurement, or the plaintext to be encrypted, is a vector with 1024 elements, all of which are integers. Gamma spectra measured in air always contain background counts because of ubiquitous presence of primordial radioisotopes on earth. Radiation background measured by NaI consists primarily of gamma emission of primordial radioactive isotope $^{40}$K, and the daughter products of the natural decay series of $^{238}$U, $^{232}$Th and $^{235}$U primordial isotopes [14]. Measurements made with NaI detector in air consist of one-second samplings of the nuclear background. Because of low signal to noise ratio, one-second background spectra measurement vectors do not display distinct isotopic spectral lines. If isotopic spectral lines are detectected in the measured spectrum, this is taken as an indicator of the presence of a nuclear source. An example of a typical one-second background spectrum measurement vector is shown in Fig. 1.

Fig. 1. One-second background spectrum measurement vector

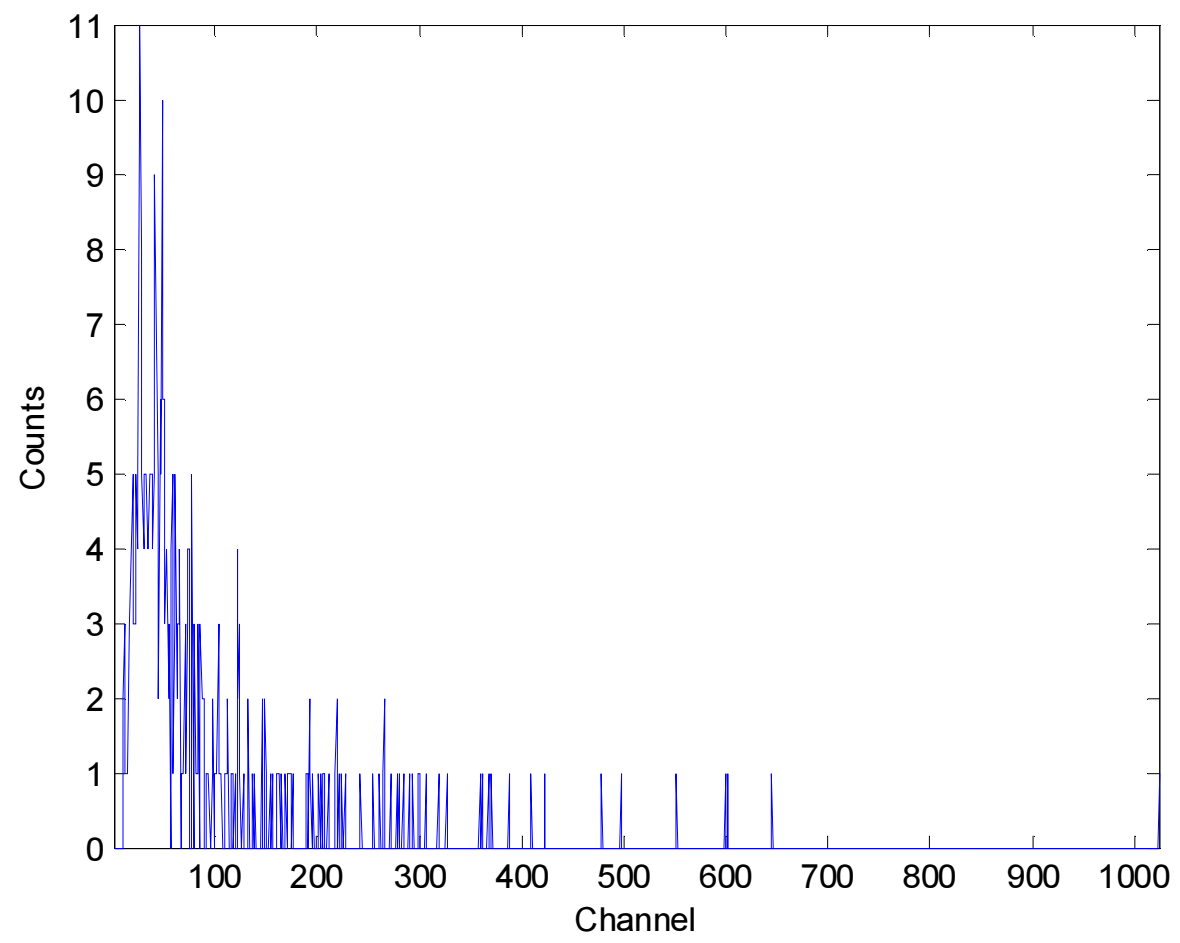

In order to demonstrate isotopic structure of urban nuclear background, Fig. 2 displays nuclear background spectrum (counts in each channel) integrated over the search time 1026 seconds. This is a log-linear plot of counts vs. energy, where the space of 1024 channels is linearly mapped to the energy space of 0 to 3000KeV. The spectrum in Fig. 2 shows recognizable isotopic lines of $^{40}$K, $^{232}$Th, and daughter products of $^{235}$U/ $^{238}$U.

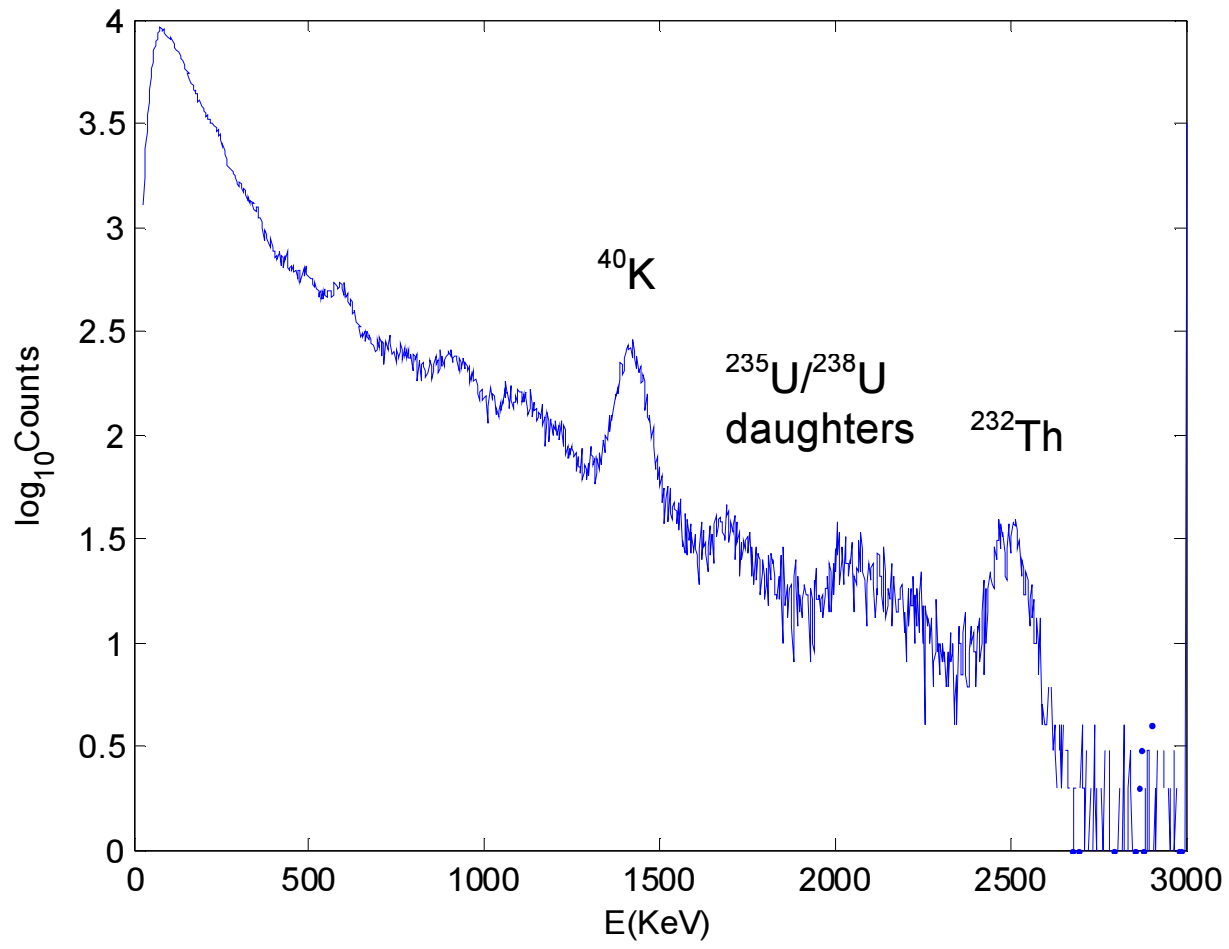

Fig. 2. Log-linear plot of time integrated counts vs. energy of the natural background nuclear spectrum.

## IV. Numerical Experiments with Compressive Sensing

### *A. Signal Compression and Reconstruction with Compressive Sensing*

In order to evaluate CS-based data compression and encryption concepts, we have conducted numerical experiments using L1-MAGIC software package operating in MATLAB environment [15]. One can observe in Fig. 1 that one-second spectrum is sparse, i.e., majority of channels record no counts. Therefore, the sparset $\mathbf{\Phi}$ for a one-second spectrum consists of 1024 spectral channels. Given that the original spectrum (in Fig. 1) $\boldsymbol{x}$ is a vector of 1024 elements, we chose the size of the compressed vector $\boldsymbol{y}$ to be a vector of 512. Thus we selected a projection matrix **A** to be 512x1024 elements. Entries of **A** were chosen as Gaussian random variables and the columns of were **A** were made orthonormal.

Compressed and encrypted measurement vector **y**=**Ax** (**y** is 512x1) is displayed in Fig. 4. Code runtime for producing the measurement vector **y** was substantially smaller than one second on a laptop with Windows XP.

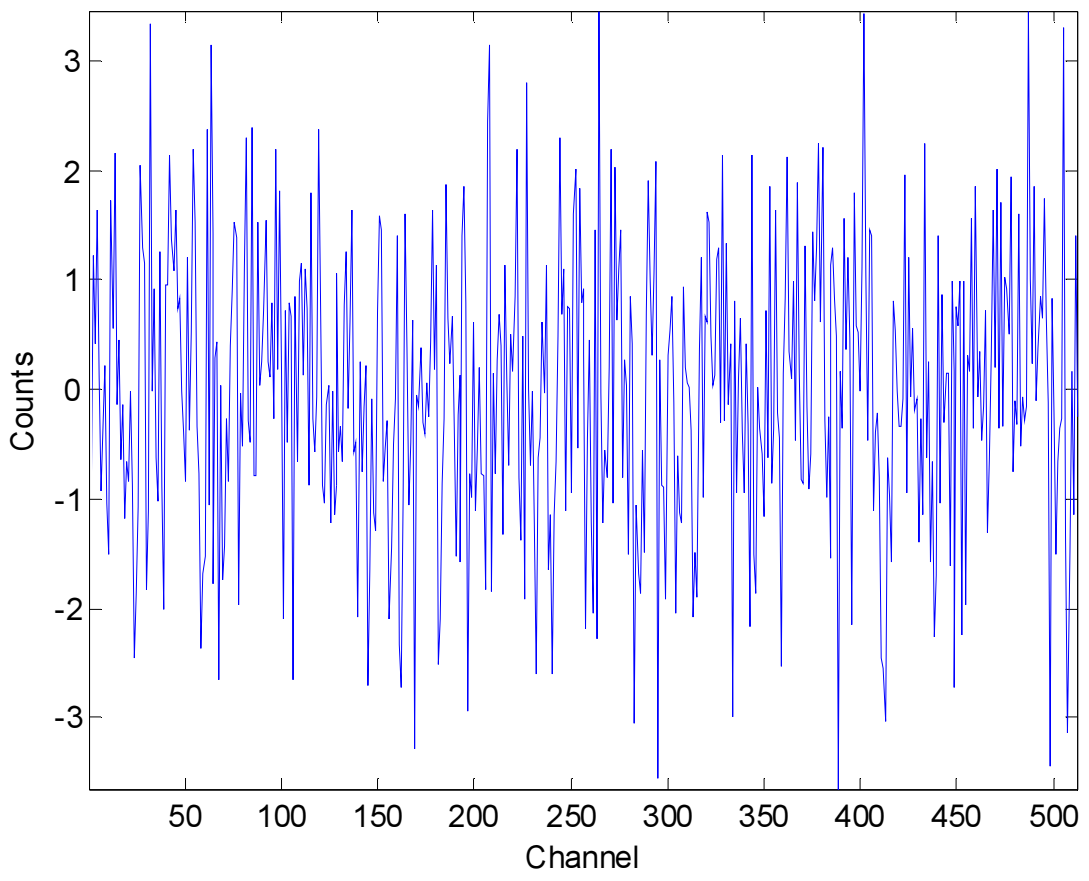

Fig. 4. Compressed and encrypted measured vector **y**=**Ax** (**y** is 512x1)

Next, the signal was reconstructed using the iterative optimization routines of L1-Magic. The iteration procedures begins with the initial guess of $\mathbf{x}_0=\mathbf{A}^T\mathbf{y}$. Reconstructed vector $x_r$ (which has the size of the original vector 1024x1) is displayed in Fig. 5. The reconstruction error was $||\mathbf{x} - \mathbf{x}_r|| \sim 10^{-4}$, which implies essentially lossless recovery of the encrypted/compressed signal. Code runtime for signal reconstruction was approximately one second on a laptop with Windows XP.

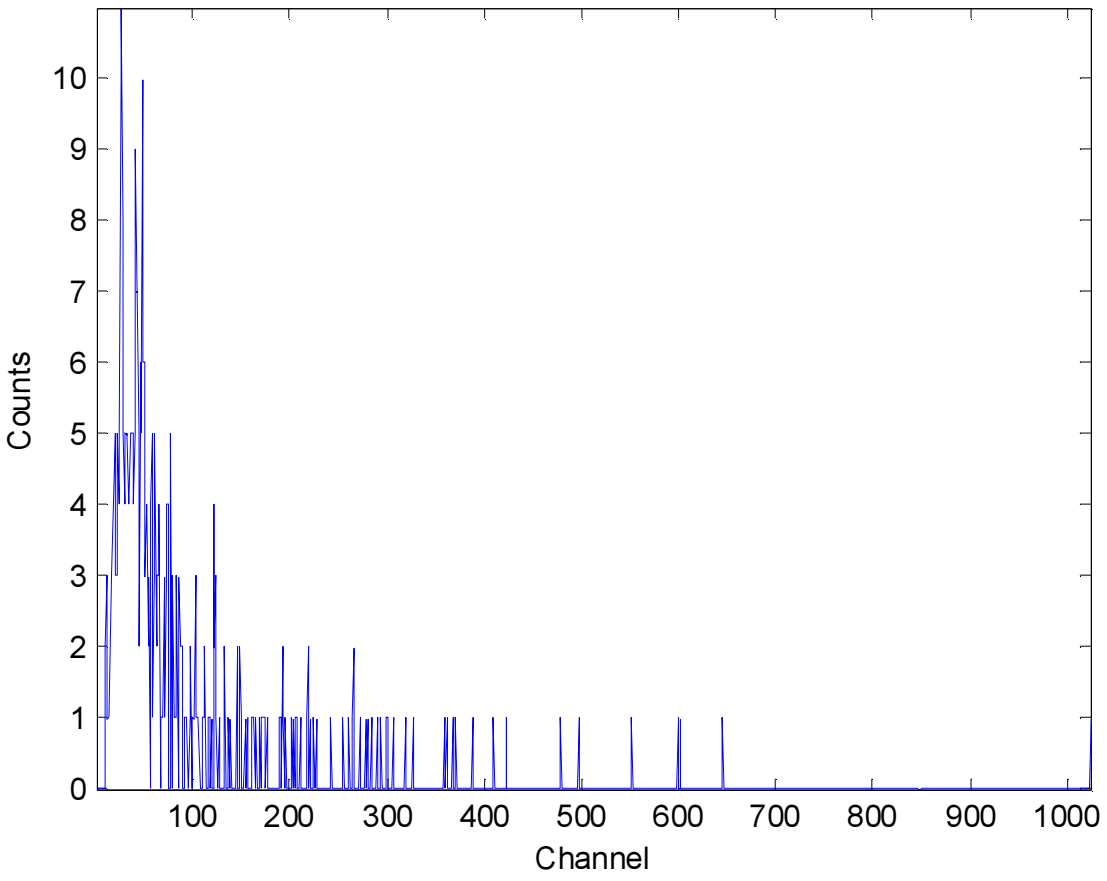

Fig. 5. Reconstructed vector $\mathbf{x}_r$ ($\mathbf{x}_r$ is 1024x1). Reconstruction error $||\mathbf{x} - \mathbf{x}_r||$ is on the order of $10^{-4}$

### *B. Simulation of Blind Reconstruction of Intercepted Signal*

In order to demonstrate encryption strength of CS, we simulated a scenario in which an eavesdropper intercepts the encrypted signal **y** in Fig. 4. The eavesdropper does not have the encryption key, so he or she attempts to perform blind reconstruction of the intercepted signal. Suppose the eavesdropped correctly guesses the size of the measurement matrix **A** (512x1024). Eavesdropper's blind reconstruction vector **z** (1024 elements) with a randomly chosen 512x1024 projection matrix is shown in Fig. 6. Clearly, eavesdropper's reconstructed signal **z** has no resemblance to the original signal **x**.

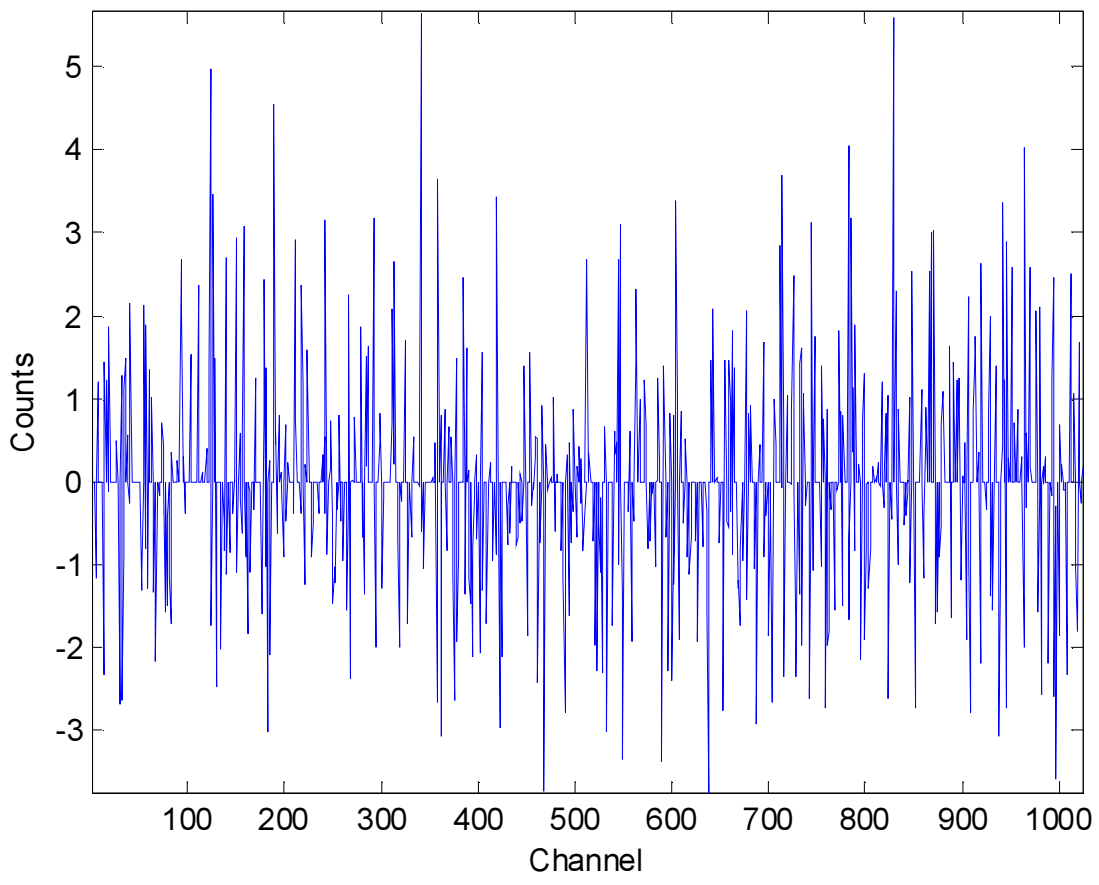

Fig. 6. Eavesdropper's blind reconstruction signal **z**

### *C. Parametric Study of Reconstruction Error*

We conducted a set of numerical experiments in order to investigate limits on data compression and reconstruction reliability with CS. Using the original signal $\boldsymbol{x}$ in Fig. 1, we have investigated the reconstruction error $||\mathbf{x}\text{-}\mathbf{x}_r||$ dependence on the compressed vector length (sampling size). In this study, length of the compressed vector $\boldsymbol{y}$ was varying from 256 to 768 elements (25% to 75% of the original $\boldsymbol{x}$). Each data point on the graph was obtained by generating projection matrices **A** of sizes from 256x1024 to 768x1024. Log-linear plot in Fig. 7 shows that there is a fairly sharp transition in reconstruction error from the value of ~10 (failure to converge) to the value of $\sim 10^{-4}$ (convergence). This transition occurs at the sampling size of ~440 (40% of the original). In addition, there are several sporadic instabilities (sampling ~ 470) which occur after the errors transitioned to small values. Limits of compression with CS, transitions from large to small errors, and numerical instabilities are currently not well understood in CS theory. Since these issues are highly pertinent to CS data compression performance, these questions will be investigated in the remainder of the project.

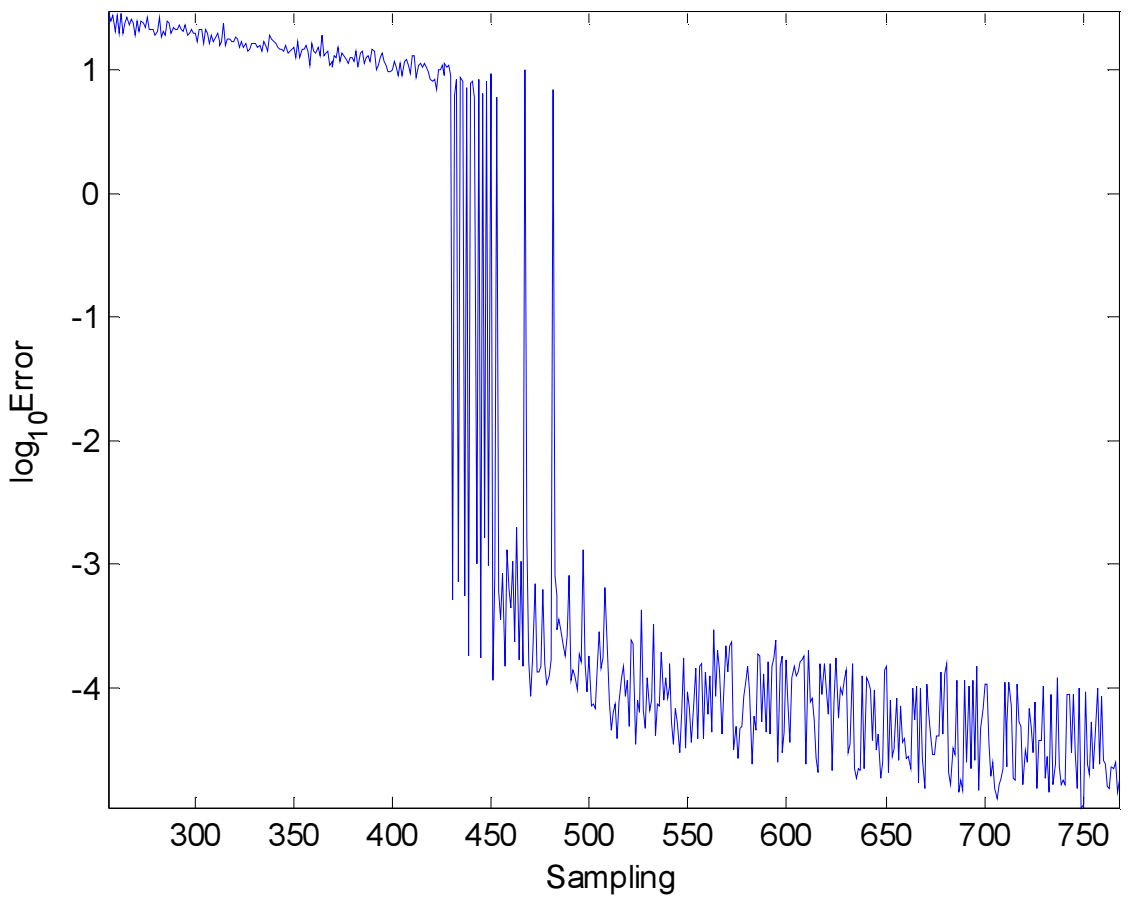

Fig. 7. Reconstruction error $||x\text{-}x_r||$ dependence on compressed vector length (sampling size)

In the next numerical experiment, we tested stability of reconstruction for the same size compression (measurement sampling). We chose the sampling value of 512, which, according to Fig. 7, would yield a small error in reconstruction. We conducted a set of 500 repeated trials in which different 512x1024 projection matrices $A$ were generated to perform CS coding and decoding of the original signal $x$ in Fig. 1. Log-linear plot of the reconstruction error $||x-x_r||$ in Fig. 8 shows that the reconstruction converges in each trial.

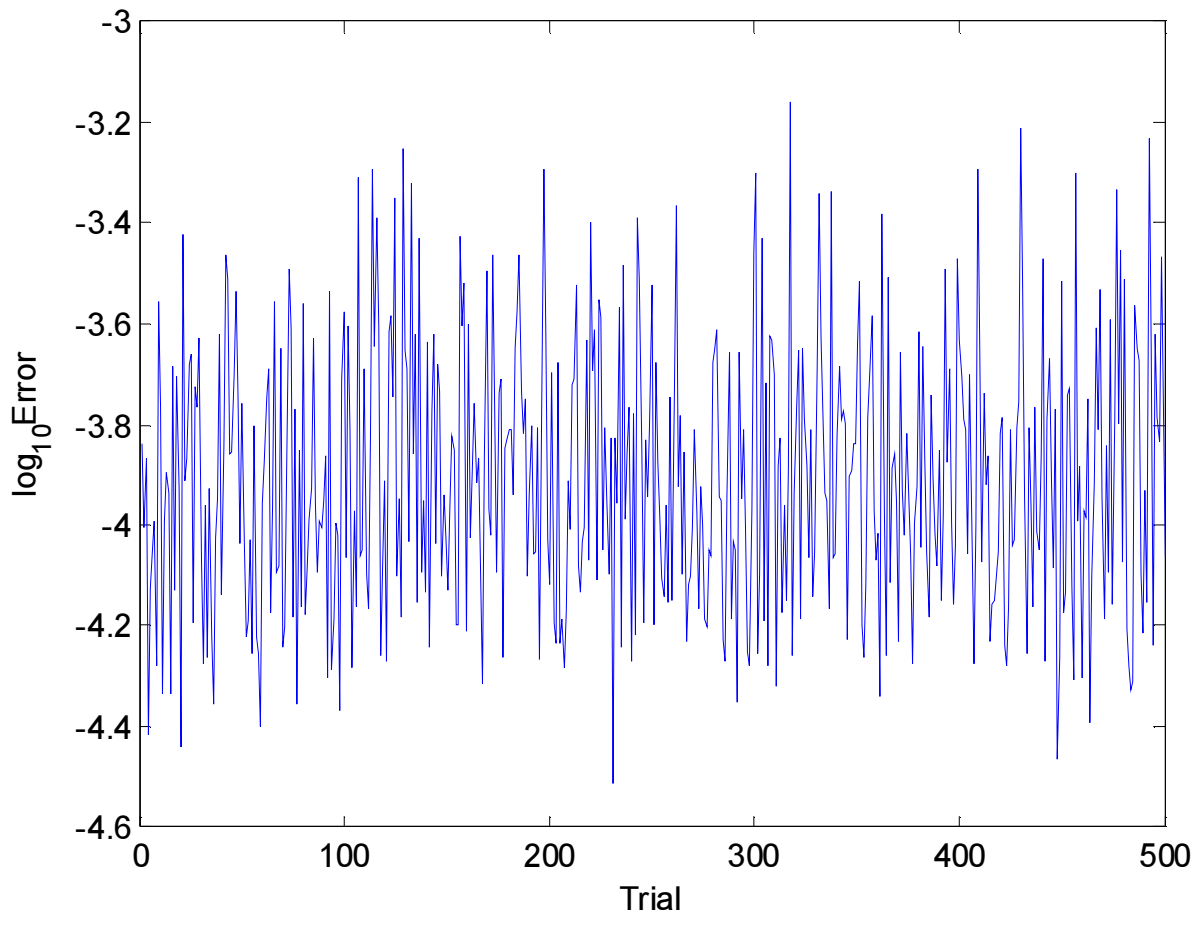

Fig. 8. Reconstruction error $||\mathbf{x}-\mathbf{x}_r||$ in 500 repeated trials with different 512x1024 projections **A**.

*D. Simulation of Coding and Decoding Operations on a Large Set of Spectra*

Next, we applied CS coding and decoding to the set of 600 one-second gamma spectra obtained in nuclear search. Integral of this data over time is displayed in Fig. 2. Using the same projection matrix **A** of size 512x1024 (i.e. compressed vector **y** is 512x1), we encoded and decoded the one second-spectra, Log-linear plot of the reconstruction error $||\mathbf{x}-\mathbf{x}_r||$ for each of 600 one-second spectra is shown in Fig. 9. Although the majority of reconstructions converge, a few reconstructions fail. Reasons for these failures (either stochastic or deterministic) will be investigated in the future work.

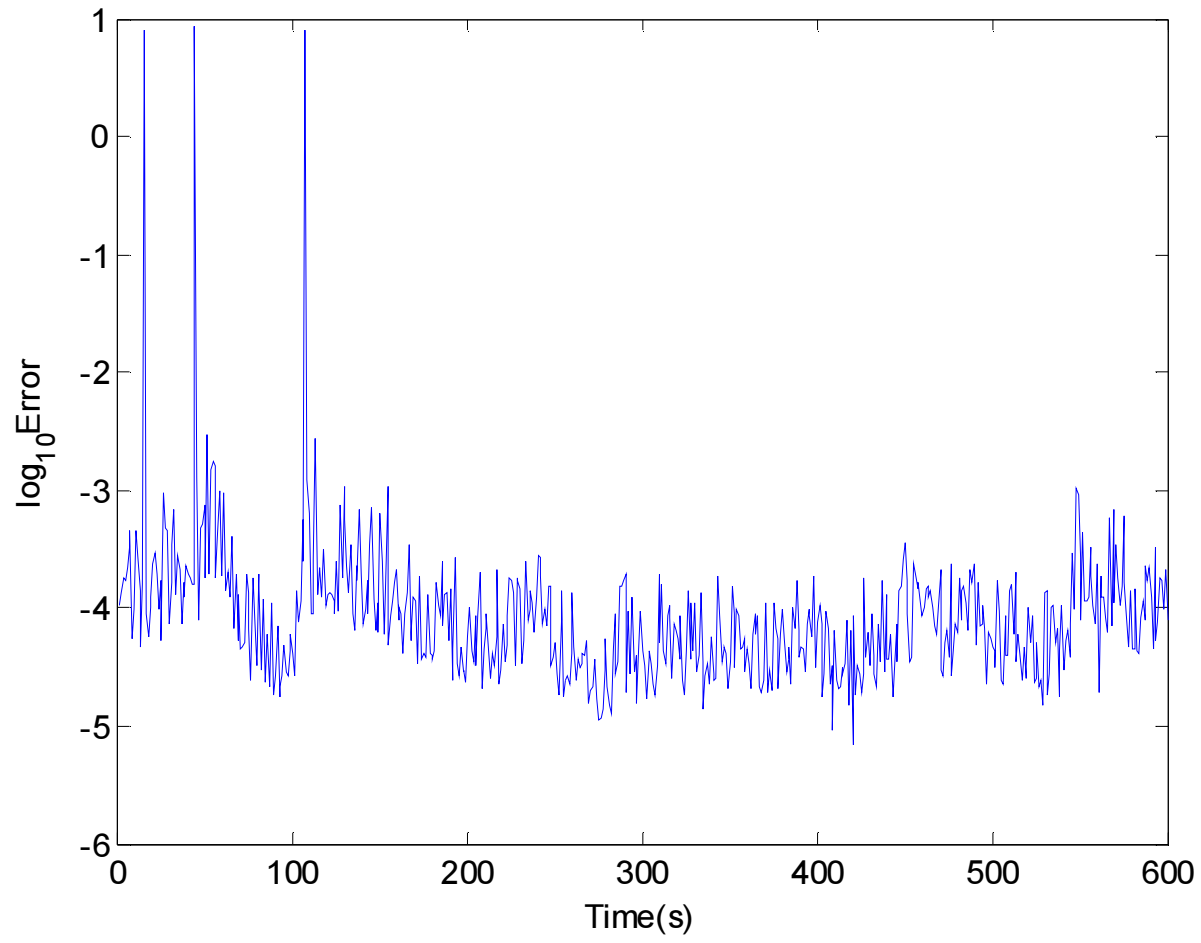

Fig. 9. Reconstruction error $||x-x_r||$ for each of the 600 one-second spectra

To demonstrate the validity of CS-based approach, we have summed up reconstructed spectra the errors for which are shown in Fig. 9. Log-linear plot of the integrated reconstructed spectrum is shown in Fig. 10. The integrated reconstructed spectrum is very similar to the integrated background spectrum in Fig. 2. Spectral peaks corresponding to the isotopic lines of $^{40}$K, $^{232}$Th and the daughter products of $^{235}$U/ $^{238}$U are clearly visible in Fig. 10. In search applications, processing of nuclear spectra is usually automated by using radioisotope identifier algorithms (RIID). In the remainder of this project, we will validate the application of CS-based encryption/decryption in nuclear search by comparing the performance of an RIID on the original and reconstructed spectra.

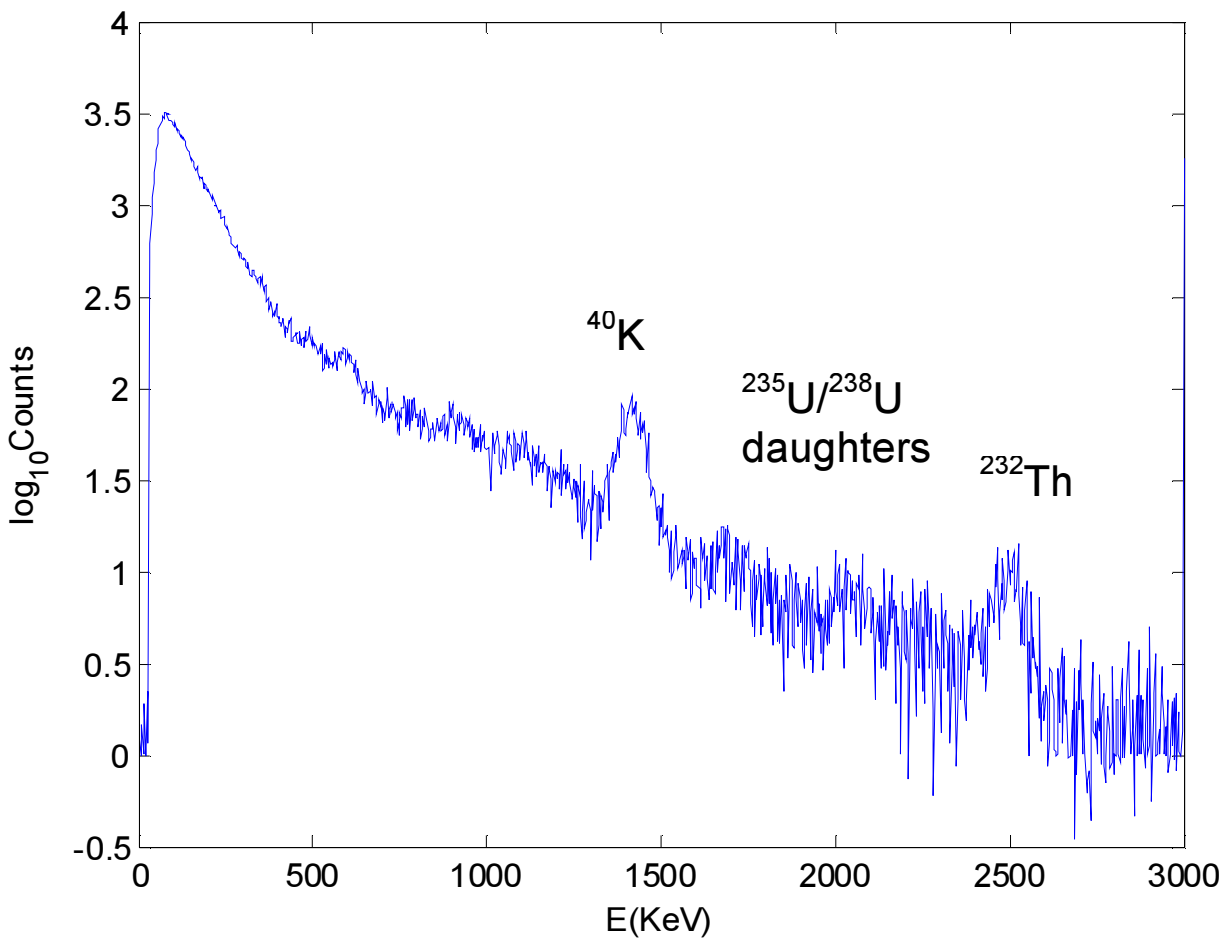

Fig. 10. Log-linear integrated reconstructed spectrum displaying characteristic isotopic spectral lines.

## V. Conclusion

In this paper, we have investigated the performance of compressive sensing (CS) computational technique in real-time compression and encryption of nuclear search spectra. Results of our numerical experiments have shown that CS provides computationally efficient compression and encryption of nuclear spectra. Future work on this project will investigate limits of signal compression with CS, sudden transitions from large to small errors in signal reconstruction, and numerical instabilities in reconstruction, and reasons for reconstruction convergence failure (either stochastic or deterministic). We will also benchmark the performance of CS-based methods against existing state-of-the-art compression and encryption methods

## Acknowledgment

This work was supported by the Argonne National Laboratory LDRD 2011-027-N0 "Data Compression and Encryption for Cybersecurity Applications."

**Alexander Heifetz** is an Electrical Engineer with the Nuclear Engineering Division at Argonne National Laboratory.
**Sasan Bakhtiari** is a Principle Electrical Engineer with the Nuclear Engineering Division at Argonne National Laboratory.
**A.C. (Paul) Raptis** is Senior Electrical Engineer with the Nuclear Engineering Division at Argonne National Laboratory